\documentclass[aps,prb, amsmath,amssymb,showpacs,superscriptaddress,twocolumn]{revtex4}
\usepackage{graphicx}
\usepackage[usenames, dvipsnames]{xcolor}
\usepackage{subfigure}
\usepackage{hyperref,hypcap}
\usepackage{braket}
\usepackage{amsmath}
\usepackage{bm}
\usepackage{cancel}
\usepackage[normalem]{ulem}

\begin{document}
\title{Polariton Supercurrent Generation in Unipolar Electro-optic Devices}
\author{Ming Xie}
\affiliation{Department of Physics, The University of Texas at Austin, Austin, TX 78712, USA}
\author{David Snoke} 
\affiliation{Department of Physics and Astronomy, University of Pittsburgh, Pittsburgh, PA 15260, USA} 
\author{A. H. MacDonald}
\affiliation{Department of Physics, The University of Texas at Austin, Austin, TX 78712, USA}

\date{\today}

\begin{abstract}
We describe a mechanism by which an electrical bias voltage applied across a unipolar 
semiconductor quantum well can drive an exciton or polariton supercurrent.  
The mechanism depends on the properties of electronic quasiparticles in quantum wells
or two-dimensional materials that are dressed by interactions with the
coherent exciton field of an exciton condensate or the coherent exciton and photon  
fields of a polariton condensate, and on approximate conservation laws.
We propose experiments that can be performed to 
realize this new light-matter coupling effect, and discuss possible applications.   
\end{abstract}

\pacs{71.35.-y, 73.21.-b, 71.36.+c}

\maketitle

\section{Introduction}
This work is motivated by remarkable progress over the past decade in realizing and 
exploring the physics of dynamical steady states of coupled light and matter in which 
two-dimensional exciton-polaritons, bosonic states that are coherent mixtures of vertical 
cavity photons and quantum-well excitons, condense into states with macroscopic
coherence.\cite{Kasprzak2006, Balili2007, Deng2010, Carusotto2013, Byrnes2014, Liew2011, Sanvitto2016, Sun2017}
The achievement of exciton-polariton condensation can be viewed as
an end point of the long quest for exciton condensation in semiconductors \cite{Koch2006, Snoke2002, Blatt1962} containing steady
state populations of conduction band electrons and valence band holes.  From this point 
of view, the role of the resonant coupling to vertical cavity photons, which turns excitons into 
exciton-polaritons, is to provide the bosonic particles with phase stiffness via the small 
polariton mass that is  sufficient to combat disorder that would otherwise localize the excitons.
In typical systems that steady-state polariton population is provided by 
optically exciting a population of non-resonant excitons that can be scattered into the condensate 
at a rate that balances the rate at which polariton population is lost by cavity-photon 
leakage.  For many purposes the polariton system environment can be approximated 
as a thermal reservoir.  The polariton condensate can also be supported electrically 
by applying a bias voltage across a bipolar system so that the current is carried through the 
system by injecting conduction-band electrons at an n-contact and valence-band holes at a 
p-contact.\cite{Schneider2013, Bhattacharya2013, Bhattacharya2014} 

In this article we describe a mechanism for electrical manipulation 
of exciton-polariton condensates that can act even in unipolar systems.  
It is based on the properties of electronic quasiparticles in quantum wells
or two-dimensional materials that are dressed by interactions with the
coherent exciton field of an exciton condensate, or the coherent exciton and photon  
fields of a polariton condensate.  It relies on approximate conservation of total exciton number in 
an exciton condensate or on approximate conservation of the sum of total  
photon and exciton numbers in polariton condensates.  Because it generates excitons or polaritons where the particle 
current exits the coherent cavity region and absorbs them when it enters, it does not yield a 
global exciton or exciton-polariton generation rate and thus it cannot on its own support a finite bosonic 
particle population.  It can however generate a supercurrent in the exciton or polariton 
condensate that flows between source and drain regions.  

Our article is organized as follows.  In Section II we explain the basic mechanism, which 
applies to both exciton and exciton-polariton condensate cases.  Throughout this article 
we focus on the exciton-polariton case because it can be reliably realized 
experimentally.  The basic idea is that when electronic quasiparticles are dressed by their 
interaction with an exciton-polariton condensate, they do not separately conserve 
conduction and valence band currents.  Conservation is recovered by compensating boson generation 
and absorption contributions to the condensate equation of motion that drive the condensate 
into a state that carries a super-current.  This effect is partly analogous to supercurrent generation 
at normal-superdonducting interfaces, and to spin-transfer torques in ferromagnets, but has 
some surprising wrinkles. In particular the direction of the exciton-polariton supercurrent is opposite 
to the direction of charged particle flow. The transfer process between the quasiparticles and the 
condensate in general leads to source terms for different components of the exciton-polariton condensate, in 
a ratio that depends on the strength of the coupling of the quasiparticles to that 
component of the condensate.  (In practice the quasiparticle-condensate transfer process
mainly produces a source term for the exciton portion of the condensate. )  In Section III 
we explain how separate transfer processes, proportional to exciton-photon Rabi coupling strengths, 
rearrange the composition of the current on a longer length scale 
so that it is ultimately carried by the lower-polariton condensate.
In Section IV we examine the possibility of identifying these transfer effects 
experimentally by measuring the finite transverse momentum of the lower-polariton condensate 
when it carries a supercurrent.  Finally in Section V we summarize our results, present conclusions, and 
speculate on possible applications.

\section{Electrically Driven Exciton Supercurrent}
\label{conversion}
\begin{figure}[t]
	\includegraphics[width=0.95\columnwidth]{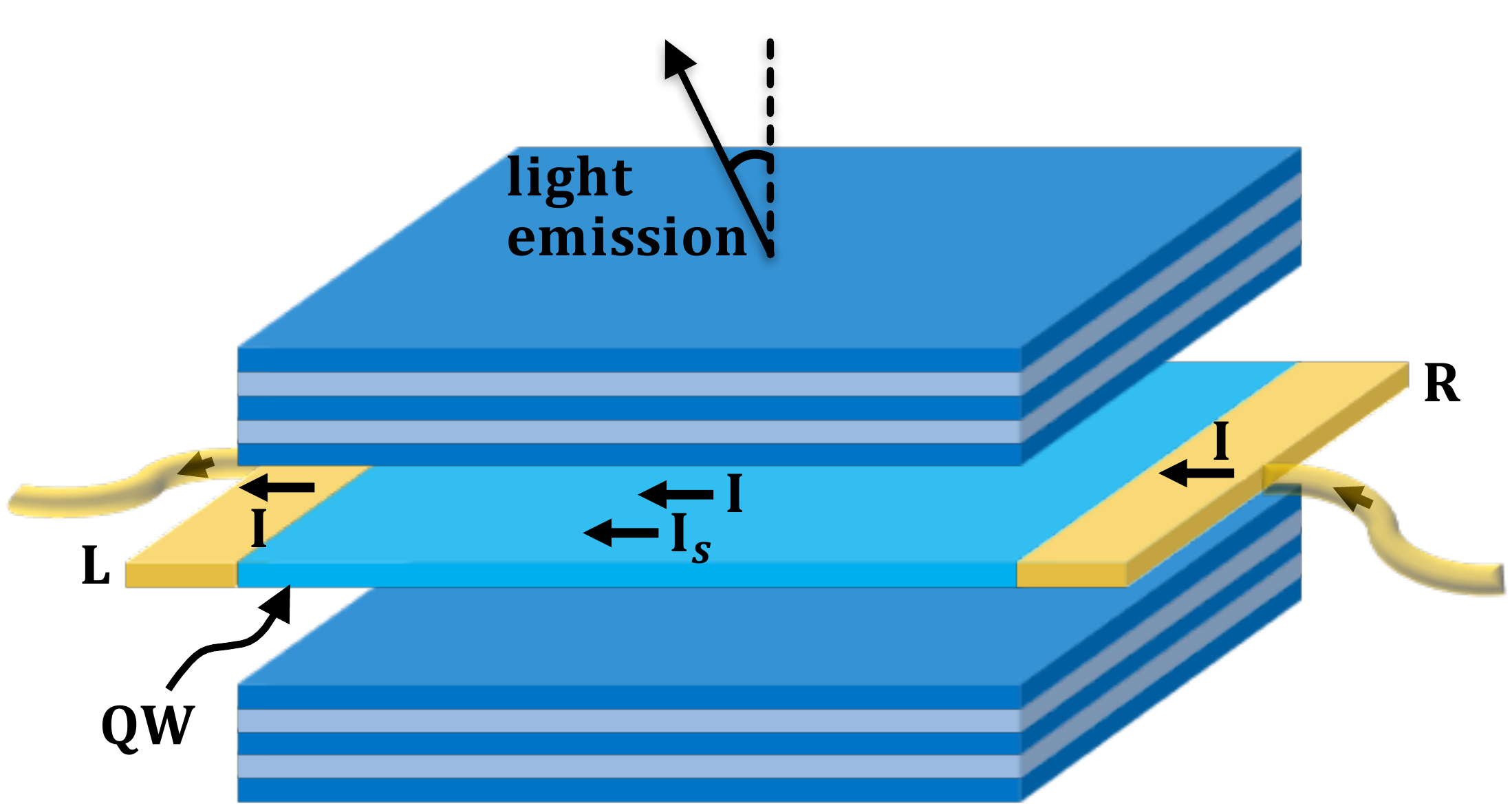}
	\caption{(Color online) Schematic representation of the planar semiconductor microcavity with contacts made to the quantum well.}
	\label{Fig:Structure}
\end{figure}

We illustrate the quasiparticle-condensate transfer effect by considering
a planar semiconductor microcavity with electrical contact established  
to an embedded quantum well, as illustrated schematically in Fig.~\ref{Fig:Structure}.
For definiteness we assume that the contacts are $n$-type and that the quantum well is intrinsic,
so that the electrical system is a lateral $n$-$i$-$n$ homojunction.
As we explain below, this geometry does not\cite{footnote1} support steady-state net 
electrically-controlled exciton or polariton generation.  
We therefore assume that a non-resonant optical pump is also present to maintain 
a finite steady-state polariton population in the cavity region.
We are interested in the regime in which the pumping power exceeds the  
threshold for polariton condensate formation, and a bias voltage is applied 
between the electrical contacts.  

The influence of electrical transport on the polariton condensate can be understood in a two-step model.
First, electric charge is injected from the left ($L$) contact, drained to the right ($R$) contact 
and transported through the cavity region by fermionic 
quasiparticles that carry an electrical charge and are dressed by the condensate.  
During this process, a neutral exciton supercurrent is induced in response to the charge current
via a mechanism explained below.  Second, the exciton supercurrent is converted to a lower polariton 
supercurrent.  During this latter process, part of the exciton condensate momentum is 
transferred to the photonic component.
This two step model is justified when the exciton-photon conversion length scale is much longer than the length scale over which the exciton supercurrent is generated.  

We describe both the quantum well and the contacts with a two-band semiconductor model.
The contacts, which are located outside of the cavity, are not dressed by the photon field, but 
the bare conduction and valence band states of the quantum well in the cavity region
are coupled by both the coherent photon and the coherent exciton field. 
For the illustrative calculations we perform below, we assume that the coherence amplitude rises steeply at 
the $n$-$i$ interface.  The mean field Hamiltonian in the coherent region in the 
bare conduction-valence basis is 

\begin{align}
H= 
\left(
   \begin{matrix}
       H_0 &  g\psi(\mathbf{r})+\Delta(\mathbf{r})   \\
       g\psi(\mathbf{r})+\Delta(\mathbf{r})         &  -H_0
   \end{matrix}
   \right)
\label{ham}
\end{align}
where $H_0= -\hbar^2{\nabla}^2/2m+ \delta/2$ and ${\nabla}$ is the 2D gradient operator. 
For simplicity, we have taken the effective masses $m$ of the two bands to be identical and isotropic. 
We further simplify the problem by assuming that the system is translationally invariant
in the transverse $(y)$ direction and neglecting electrostatic band bending effects.  

The Hamiltonian (\ref{ham}) is written in a rotating-wave frame in which the condensate 
energy is effectively absorbed into the energy of electrons, so that the actual bandgap $E_g$ is replaced by a 
reduced gap $\delta \equiv E_g - \hbar\omega$ where $\hbar\omega$ is the chemical potential of the condensate.
For typical polariton condensates $\delta$ is of the same order\cite{Byrnes2010,Kamide2010,Fei2016}
as the two-dimensional exciton binding energy.
In the two band model rotating and fixed ($\text{f}$) representation wavefunctions are related by
\begin{align}
\Psi(\mathbf{r},t) = \hat{U}^{-1}(t) \Psi_\text{f}(\mathbf{r},t)
\end{align}
where
$\hat{U}(t) = \exp(-\frac{i}{2}\omega t\sigma_z)$
and $\sigma_z$ is the band Pauli-matrix. 
Here $\Psi(\mathbf{r},t) = \left(u(\mathbf{r},t),v(\mathbf{r},t)\right)^T$ is a two-component vector
whose components are the conduction and valence components of the quasiparticle wavefunction.

The two terms in the off-diagonal component of the Hamiltonian 
originate from electron-photon coupling and 
from Coulomb interactions\cite{Byrnes2010,Kamide2010,Fei2016} respectively, and are 
time-independent in the rotating representation.  
In the electron-photon coupling term $g$ is the electric dipole coupling 
constant and $\psi=\left<\hat{a}\right>$
is the coherent photon field, {\it i.e.} the expectation value of the photon annihilation 
operator $\hat{a}$ in the coherent photon state.
$\Delta$ is a mean field describing coherence induced by interband exchange 
interactions, which are also the origin of the attractive force that binds electrons and holes
in an isolated exciton. Both $g\psi$ and $\Delta$ are chosen to be real and constant across the cavity region$(0<x<L)$.
In the left $(x<0)$ and the right $(x>L)$ contact $g\psi = 0$ and $\Delta = 0$.
The matter-photon transfer effect we describe below occurs because of spatial variation in the 
inter-band coupling terms in (\ref{ham}).   

\begin{figure}[t]
	\includegraphics[width=0.95\columnwidth]{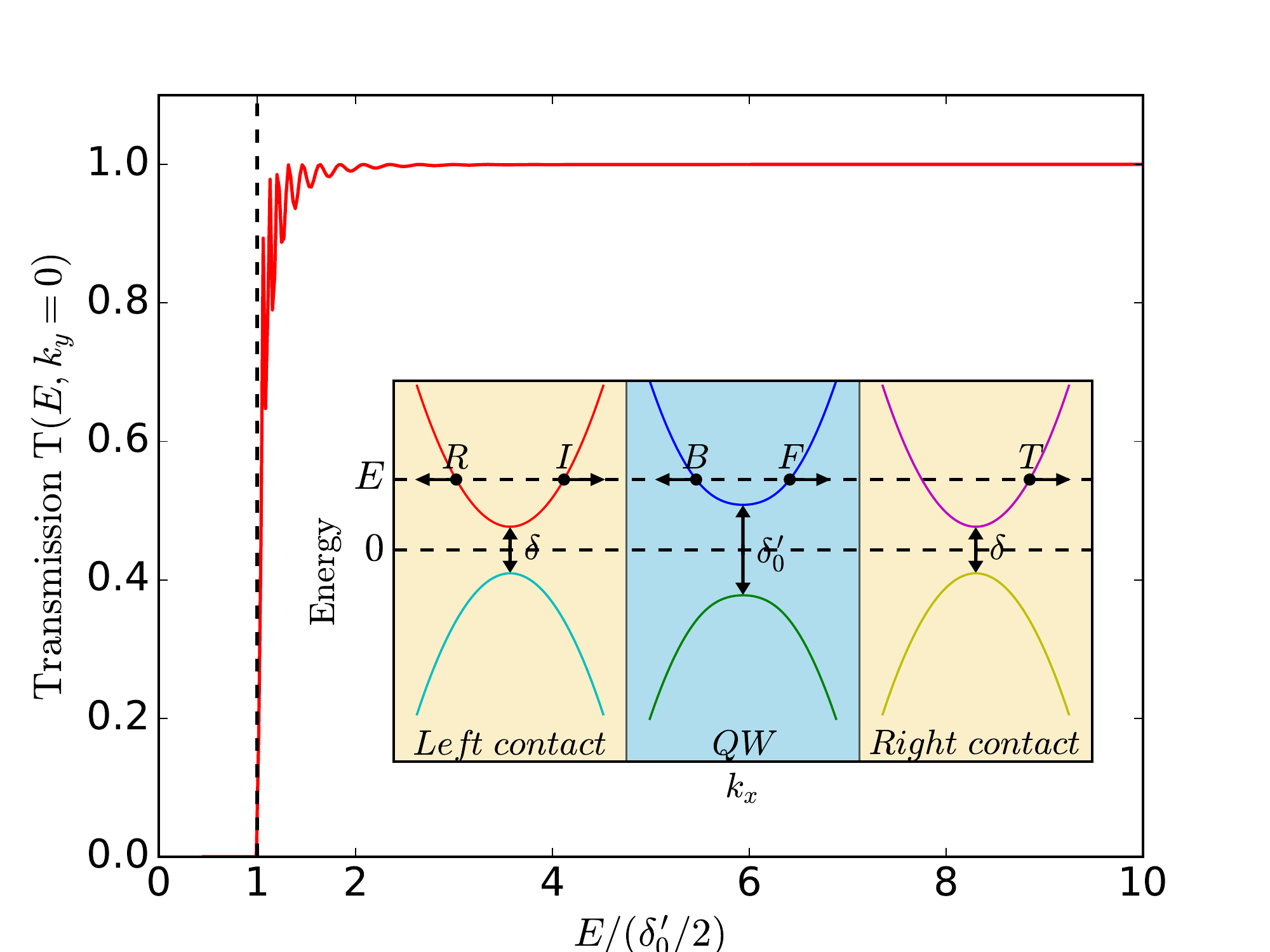}
	\caption{(color online) Energy dependence of the transmission probability for normal incidence $k_y=0$. The transition probability decays exponentially with channel length for 
	$|E| <\delta'_0/2$, where $\delta'_0$ is the renormalized quasiparticle gap.
	 Inset: Model energy dispersions in different regions for 
	 $\delta=10$meV and $g\psi+\Delta=10$meV.}
	\label{Fig:transmission}
\end{figure}

In the rotating representation the quasiparticle Hamiltonian is time independent and its wavefunction $\Psi(\mathbf{r},t)$ satisfies a time-independent Schr\"{o}dinger equation.  The quasiparticle-condensate transfer effect is most simply
illustrated by assuming ballistic transport.  We therefore postpone a discussion of scattering effects to the 
discussion section.  Due to translational symmetry in the $y$ direction, 
the transverse momentum $k_y$ is a good quantum number. 
For each $k_y$ the $x$-dependent factor in the wavefunction satisfies a one-dimensional Schr\"{o}dinger equation
with $H_0 \to H^\text{eff}_0(k_y)=-\hbar^2\nabla_x^2/2m+ \delta^\text{eff}_{k_y}/2$, where we defined 
an effective bandgap $\delta^\text{eff}_{k_y}=\delta+\hbar^2k_y^2/m$.
In the cavity region, the bulk quasiparticle energies are 
$E^\pm_\mathbf{k} = \pm E_{\mathbf{k}}\equiv \pm\sqrt{\xi_\mathbf{k}^2+(g\psi+\Delta)^2},$
where
$\xi_\mathbf{k} = \hbar^2k_x^2/2m+\delta^\text{eff}_{k_y}/2$, 
and the quasiparticle wavefunction
has mixed conduction and valence band character.
It is useful to define a $k_y$-dependent quasiparticle bandgap 
$\delta'_{k_y}\equiv \sqrt{(\delta^{\mathrm{eff}}_{k_y})^2+4(g\psi+\Delta)^2}$.
Fig.~\ref{Fig:transmission} plots the transmission probability for a $k_y=0$ electron incident ($I$) from
the left contact as a function of energy $E$.  For $E>\delta'_{0}/2$ propagating modes are present 
in the cavity region and the transmission probabilities between source and drain contacts are large.
Because the channel length exceeds microscopic length scales,
the transmission probability jumps sharply from near $0$ to near $1$ at the band edge.
(The oscillatory behavior in Fig.~\ref{Fig:transmission} is a consequence of multiple reflection between source/bulk and 
bulk/drain interfaces.) 

\begin{figure}[t]
	\includegraphics[width=0.95\columnwidth]{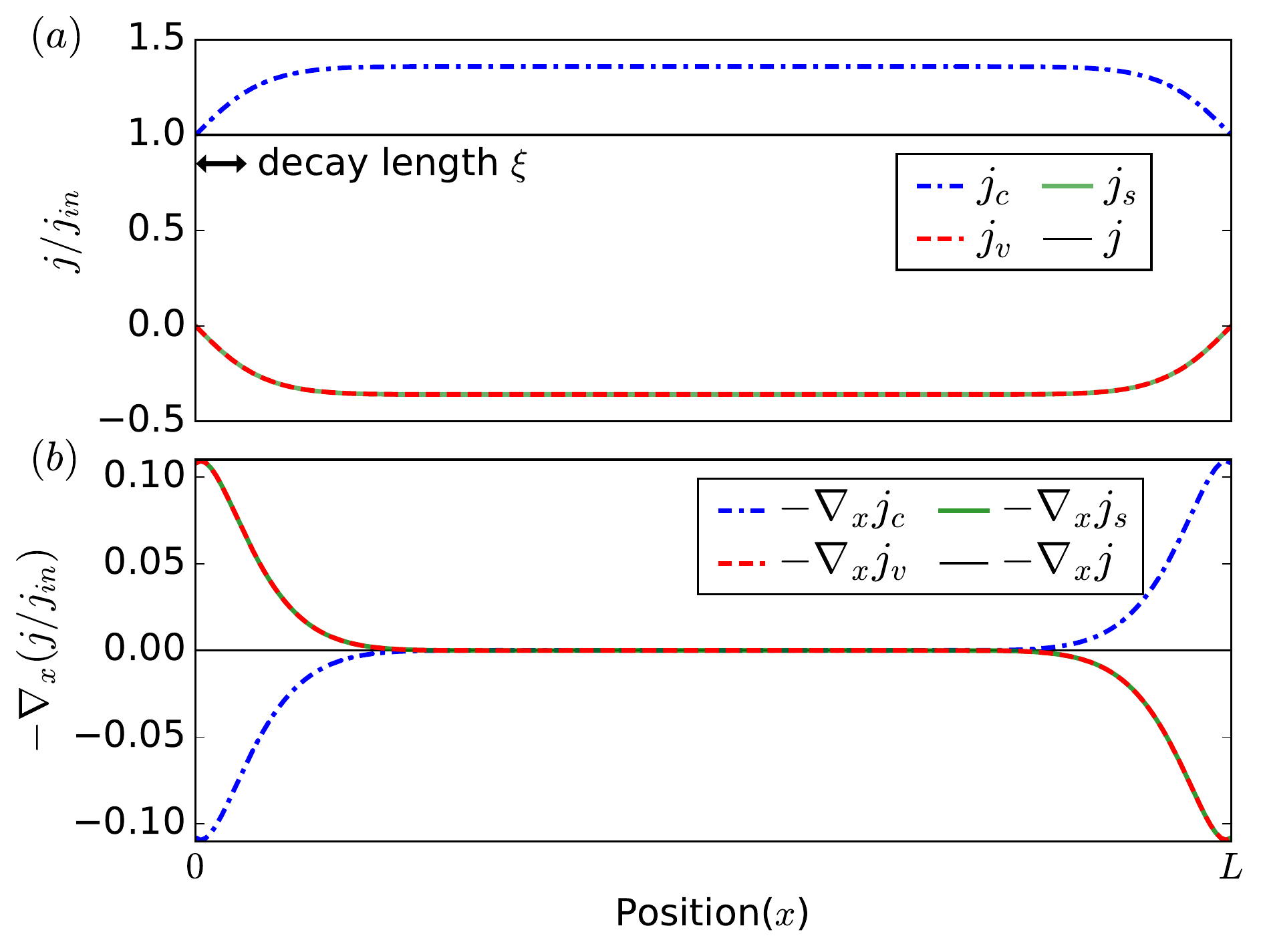}
	\caption{(color online) Spatial dependence of (a) normalized conduction and valence band 
	partial probability currents and (b) divergence of the probability currents inside the QW for normal incidence $k_y=0$ with $E = 0.55\delta'_0$. $j_\text{in}$ is the current
	incident from the left contact. 
	For the chosen parameters, the decay length $\xi \approx 1.8$ nm is much smaller 
	than the channel length $L$.}
	\label{Fig:Spatial}
\end{figure}

Because the condensate is charge neutral, electric current in the QW is carried by quasiparticle excitations only. 
For energy $E > \delta'_{k_y}/2$ the cavity 
region has two conduction-band dominated propagating Schr\"{o}dinger equation solutions with 
real wavevectors $\pm k_x^+$, and two valence-band dominated evanescent Schr\"{o}dinger equation solutions
with imaginary wavevectors $\pm k_x^-$.  
Here $k_x^\pm=2m\sqrt{-\delta^\text{eff}_{k_y}/2 \pm (E^2-(g\psi+\Delta)^2)^{1/2}}/\hbar^2$. 
We solve the elastic scattering problem by matching the wavefunction and its derivative at the two interfaces
and conserving transverse momentum\cite{Blonder1982,Rontani2005}.
Quasiparticle flow is characterized by the counter flow of its conduction and valence component.
From the mean field equation, we obtain the following coupled continuity relations
\begin{align}
\frac{\partial |u|^2}{\partial t}+ \nabla_x j_{c} &= \frac{2(g\psi+\Delta)}{\hbar}\text{Im}\{u^*v\} \label{eom_c}
\\
\frac{\partial |v|^2}{\partial t}+ \nabla_x j_{v} &= -\frac{2(g\psi+\Delta)}{\hbar}\text{Im}\{u^*v\} \label{eom_v}
\end{align}
where
$j_{c}= \hbar \text{Im}\{u^*\nabla_x u\}/m$ is the conduction band current and 
$j_{v}=-\hbar \text{Im}\{v^*\nabla_x v\}/m$ is the valence band current.
(We have suppressed the $(E, k_y, x)$ dependence of the wavefunctions and the probability current densities for notational convenience.)
These relations lead directly to the conservation law of particle current 
$\partial \rho/\partial t+\nabla_x j=0$,
where $\rho=|u|^2+|v|^2$ is the particle density and
$j=j_{c}+j_{v}$ the corresponding probability current density. 
In a steady state, $j$ is spatially constant as shown in Fig.~\ref{Fig:Spatial}(a). 

The key observation from Fig.~\ref{Fig:Spatial} is that because the quasiparticles 
in the coherent cavity region are coupled to the condensate, the individual currents $j_c$ and $j_v$ are not separately
conserved.  In analogy with the spin density in a spin-$\frac{1}{2}$ system, we can define a difference density
$\rho_{\mathrm{-}} \equiv |u|^2-|v|^2$ which can be interpreted as the exciton probability distribution.
The associated continuity relation has the form
$\partial \rho_\mathrm{-}/\partial t+\nabla_x j_\mathrm{-}=4(\tilde{g}\psi+\Delta)\text{Im}\{u^*v\}/\hbar$, 
where $j_-\equiv j_\text{c}-j_\text{v}$ is the corresponding electron-hole (or counter-flow) current.
The source(or drain) term on the right-hand side is proportional to the coherence amplitude and large in the 
transition region between contact and cavity (Fig.~\ref{Fig:Spatial}(b)).  
Since the microscopic model neglects processes, other than coupling to the 
cavity photon, which violate separate particle-number conservation, a compensating source or 
drain term appears in the equation of motion for the condensate and restores the symmetry.
(See Appendix A for a formal analysis.) 
These conservation terms are precisely analogous to those that produce Cooper-pair 
source and drain terms when current flows across a normal-metal/superconductor\cite{Andreev1964} interface,
and to the spin-transfer torque \cite{Takei2014} contribution to the equation of motion for a 
magnetic condensate in inhomogeneous magnetic systems.  
In both of these cases the conservation law is restored by the action of the quasiparticles on the 
collective degrees of freedom of the condensate.
In our case, because the quasiparticles interact much more strongly\cite{Fei2016}
with the excitonic part of the condensate than with the cavity photons, the quasiparticles act primarily as a 
source for the excitonic part of the condensate.  It follows that 
$\nabla_x j_\text{s}=-2(\tilde{g}\psi+\Delta)\text{Im}[u^*v]/\hbar$.
Including this superfluid component, the conservation is recovered for the total electron-hole probability 
current $j_\text{e-h}=j_-+2j_s$, where the factor 2 appears because we define $j_s$ as the pair probability current.
Fig.~\ref{Fig:Spatial}(b) shows the divergence of the probability currents along the transport direction
which represents the rate of transfer between the quasiparticles and the condensate.

As indicated in Fig.~\ref{Fig:Spatial}, the length scale over which the quasiparticle-condensate transfer 
occurs is short.  From our calculation we can associate this length with the decay length of the 
evanescent valence-band dominated 
Schr\"{o}dinger equation solutions at conduction band energies.  
Away from the interfaces, only propagating modes survive so $j_s$ saturates at its asymptotic value $j_s^\text{sat}$.
An analytical result can be obtained for $j_s^\text{sat}$ as follows. 
Consider two cross sections one deep in the left contact $x=-\infty$ and another in the asymptotic 
region of the quantum well, e.g. near $x=L/2$. 
For an electron incident at an energy $E>\delta'_{k_y}/2$, conservation laws require 
\begin{align}
j_\text{c}|_{x=-\infty}&=j_\text{c}|_{x=L/2}+j_\text{s}^\text{sat}, \\
j_\text{v}|_{x=-\infty}&=j_\text{v}|_{x=L/2}-j_\text{s}^\text{sat}.
\end{align}
For transport through conduction quasiparticle channel, 
$j_\text{c}|_{x=-\infty}=j_\text{in}=j$ and $j_\text{v}|_{x=-\infty}=0$.
Using the relation 
$(j_\text{c}/j_\text{v})_{x=L/2}=-u_\mathbf{k}^2/v_\mathbf{k}^2$, we have $j_s^\text{sat}=j_\text{v}|_{x=L/2}$
and
\begin{align}
\frac{j^\text{sat}_\text{s}}{j_\text{}}
=-\frac{v_\mathbf{k}^2}{u_\mathbf{k}^2-v_\mathbf{k}^2}
\equiv \alpha_\mathbf{k},
\label{ratio}
\end{align}
where $u^2_\mathbf{k}=(1+\xi_{\mathbf{k}}/E_{\mathbf{k}})/2$ and $v^2_\mathbf{k}=(1-\xi_{\mathbf{k}}/E_{\mathbf{k}})/2$ are the coherence factors .
Because $u^2_\mathbf{k}-v^2_\mathbf{k}>0$, the minus sign indicate the direction of supercurrent density $j_\text{s}$ is opposite to the particle current density $j_\text{}$.

\begin{figure}[h!]
	\includegraphics[width=0.95\columnwidth]{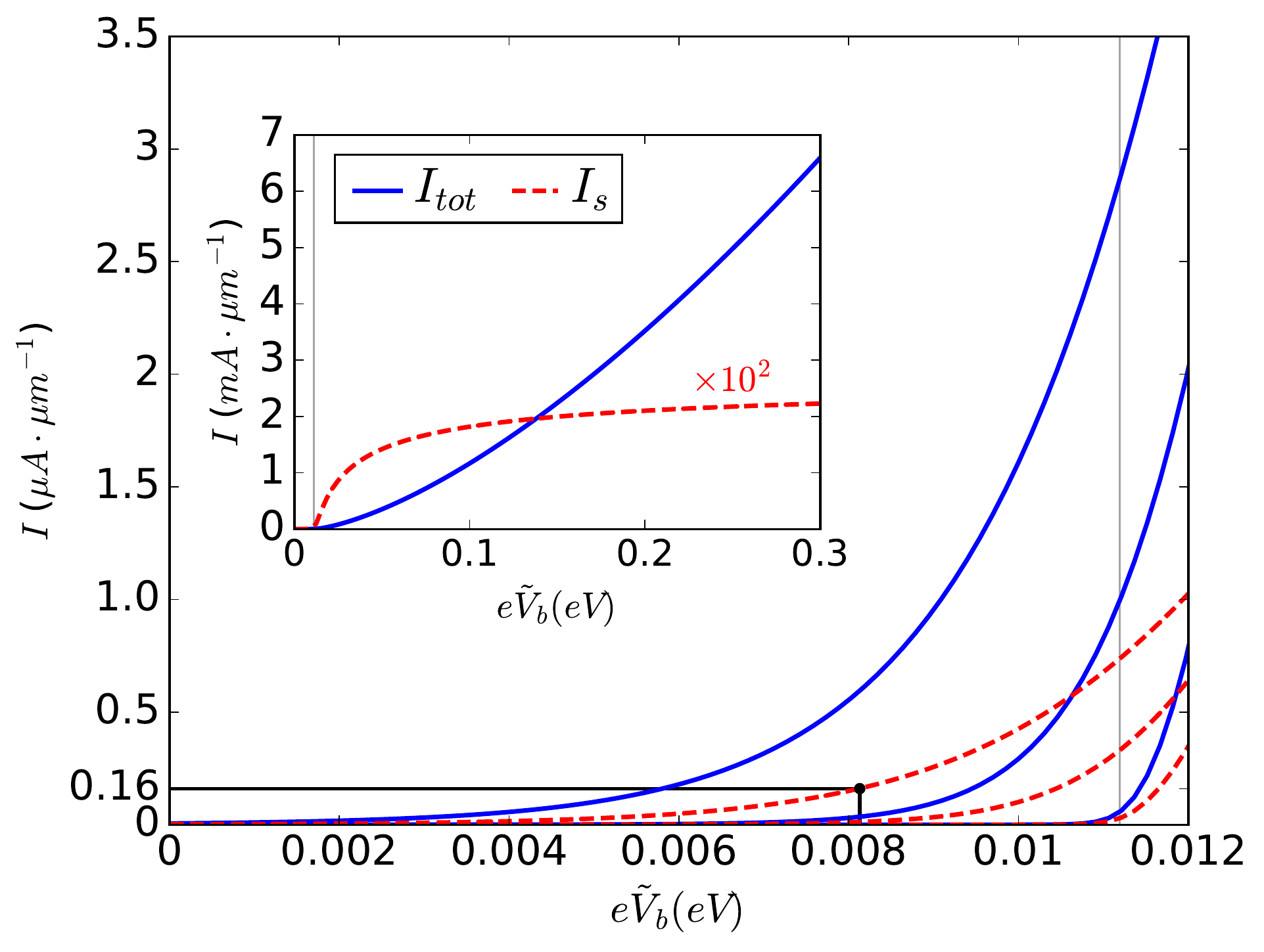}
	\caption{(Color online) Current-voltage characteristics at temperatures $T=2K,\ 10K$ and $20K$ (increasing from bottom to top.) $\tilde{V}$ is the effective bias voltage in the rotating frame. The grey line marks the threshold at the quasiparticle gap $e\tilde{V}=\delta'_0/2$. Black dot marks the minimum current $I_s^{\rm min}$ required for an observable momentum shift at $T=20K$.}
	\label{Fig:IV}
\end{figure}

\begin{figure}[h!]
	\includegraphics[width=0.75\columnwidth]{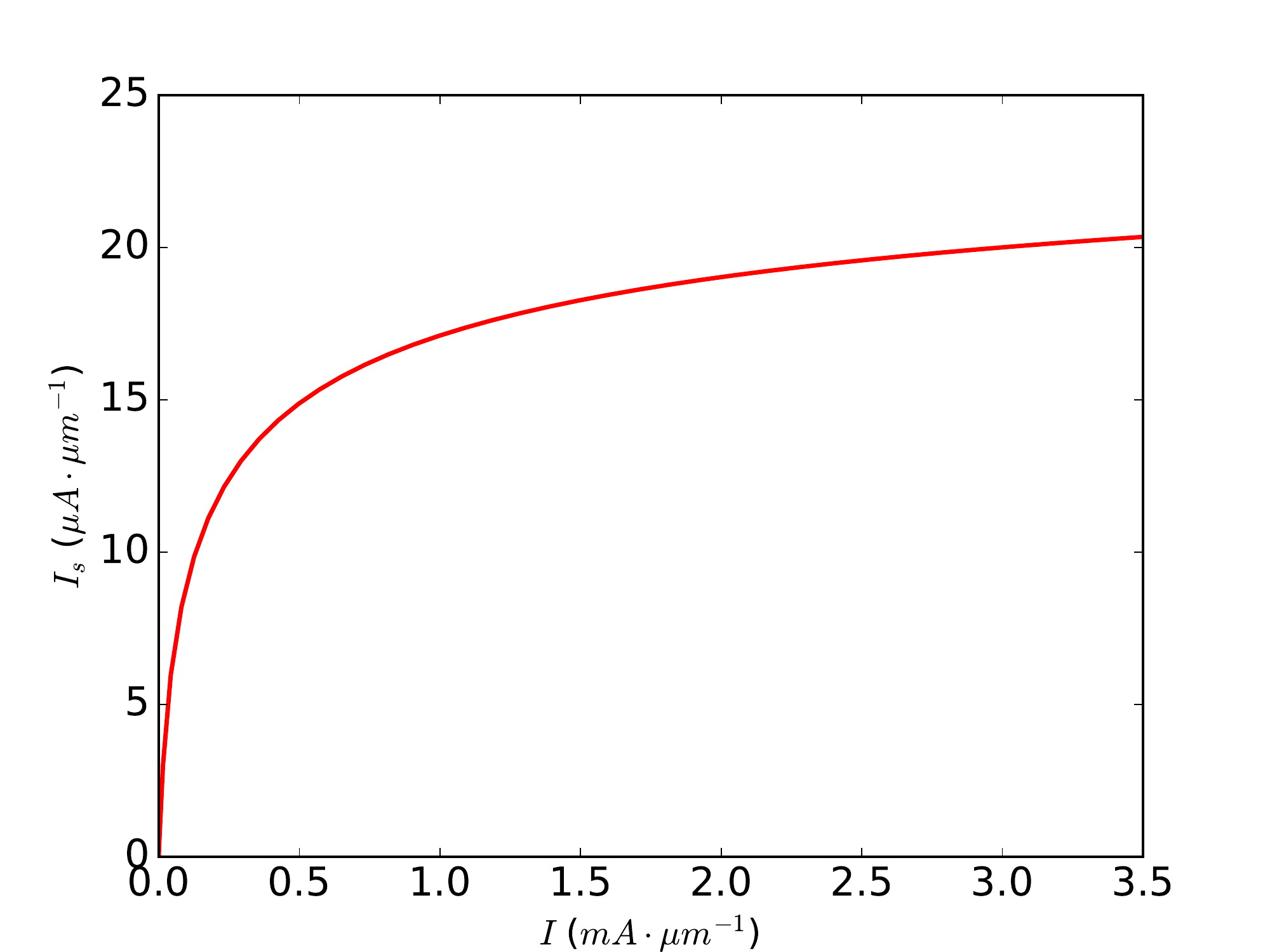}
	\caption{(Color online) Supercurrent as a function of total charge current at $T=2K$. The parameters used here are the same as in Fig.(\ref{Fig:IV})}
	\label{Fig:Is_Itot}
\end{figure}

We calculate the total charge current density in the contact using the Landauer-B\"{u}ttiker formula
\begin{align}
I_{tot}
&=\frac{-e}{hL_y}\sum_{k_y}\int T(E, k_y)\big[f(E-e\tilde{V}_b)-f(E)\big] dE
\label{i_Q}
\end{align}
where 
$e>0$ is the unit of electric charge and
$f(E)=1/[e^{-(E-\mu_R)/k_BT}]$ the Fermi distribution function.
We have assumed that the chemical potential of the right contact is fixed at the middle of the bandgap $\mu_R=0$ while $\mu_L=eV_b>0$ varies. 
The Schr\"{o}dinger equation is time-independent in the rotating frame in which the
time-dependence $e^{i\omega t}$ is gauged away. 
Provided that electrons are injected into and extracted from the 
conduction band the gauge transformation does not shift the effective bias voltage $\tilde{V}_b$.
We calculate the saturated supercurrent from Eqs.(\ref{ratio}) and Eq.(\ref{i_Q}) which yield 
\begin{align}
I_\text{s}
=\frac{e}{hL_y}\sum_{k_y}\int T(E, k_y)\alpha(E) \big[f(E-e\tilde{V}_b)-f(E)\big]dE
\label{i_s}
\end{align}
where we have arbitrarily multiplied the exciton number current by $e$ in order to 
place both currents in the same units.  Here 
$\alpha(E,k_y)=\alpha(E)$ is independent of $k_y$ and vanishes at energies far above the 
band gap.  
As a consequence, when the voltage is increased, the supercurrent will at first 
increase gradually, before saturating at higher voltages. (See the inset of Fig.~\ref{Fig:IV}.)
The supercurrent $I_s$ is plotted as a function of the total input charge current $I_{tot}$ in Fig.~\ref{Fig:Is_Itot}.
At low bias voltages, $I_s$ increases linearly with $I_{tot}$ with a slope 
proportional to the exciton density $n_{ex}$.
When the bias moves deep into the conduction band, 
$I_s$ saturates, in agree with the saturation shown in the $I-V$ curve.

\section{Exciton-Photon Transfer}

In this section, we address the length scale over which the exciton supercurrent is converted to 
a lower polariton supercurrent.  The large value of this length scale compared to the length 
scale on which the exciton supercurrent is generated justifies our simplified two-step model. 
To identify this length scale, we turn to the bosonic description of coupled 
coherent exciton $\Phi_\text{ex}$ and coherent photon $\Phi_\text{ph}$ fields.
These are described by 
the coupled linear Gross-Pitaevskii equations\cite{Pitaevskii2003, Ciuti2005}
\begin{align}
\hbar \omega \Phi_\text{ex}(\mathbf{r}) &= 
\left[-\frac{\hbar^2\nabla^2}{2m_\text{ex}}+\epsilon_\text{ex}^0\right]\Phi_\text{ex}(\mathbf{r})
+\Omega(x)\Phi_\text{ph}(\mathbf{r}),
\\
\hbar \omega \Phi_\text{ph}(\mathbf{r}) &= \left[-\frac{\hbar^2\nabla^2}{2m_\text{ph}}+\hbar\omega_0\right]\Phi_\text{ph}(\mathbf{r})
+\Omega(x)\Phi_\text{ex}(\mathbf{r}),
\end{align}
where we have neglected the weak nonlinear term that accounts for exciton-exciton interactions
and assumed that the condensate is in a quasi-equilibrium in which compensating 
pump and decay terms can be ignored.
We focus on the right boundary of the cavity  where the exciton supercurrent is generated 
and flows towards left boundary, and define the cavity region by imposing a step-like profile on the 
Rabi coupling: $\Omega(x)=\Theta(x)\Theta(L-x)\Omega_0$.
The reverse process at the left boundary region of the cavity can be considered in a similar fashion.
In the bulk, the resonant coupling between exciton and photon modes produces two new 
eigen-branches, the lower polariton(LP) and the upper polariton(UP), which 
have quadratic dispersion around zero momentum and 
effective masses $m^{-1}_\text{UP/LP} \approx m^{-1}_\text{ph}
\big[ 1 \pm (\hbar\omega_0-\epsilon^0_\text{ex})/\sqrt{(\hbar\omega_0-\epsilon^0_\text{ex})^2 + 4\Omega^2}  \,\big]/2$. 

At $x=L^{-}$ the exciton component of the condensate carries a finite supercurrent.
The condensate may also have a photon component which does not carry a current.  
When the condensate is decomposed into UP and LP components, and the condensate has formed with a 
chemical potential $\hbar\omega$ near the bottom of the LP branch, its UP component is evanescent.
Because Rabi coupling conserves total boson number while allowing 
for conversion between excitons and photons, the total current
is identical with the exciton supercurrent generated at the first step.
It follows that the exciton supercurrent generated by quasiparticle-condensate 
transfer at the boundary of the coherent region is identical to the polariton supercurrent carried in the bulk 
of the coherent region where the ratio between photon and exciton parts of the condensate 
is that of the LP mode.   
The length scale over which the LP form of the condensate is recovered 
is the decay length $\lambda_{UP}$ of the evanescent upper polariton (UP) wave which satisfies 
\begin{align}
\lambda_{UP}^{-2} =\frac{2m_\text{UP}\sqrt{(\hbar\omega_0-\epsilon^0_\text{ex})^2 + 4\Omega^2}}{\hbar^2}
\end{align}
assuming that $\hbar\omega$ is close to the LP band bottom.
Given the typical detuning $\hbar\omega_0-\epsilon^0_\text{ex} \sim 10$meV and Rabi coupling $\Omega \sim 10$meV 
values we estimate that $\lambda \sim 200$nm.
Generally $\lambda^2 /\xi^2 \approx  m_\text{0} /m_\text{ph} \sim 10^{4}$, 
provided that the renormalized gap $\delta'_0$ and the
UP-LP splitting $\sqrt{(\hbar\omega_0-\epsilon^0_\text{ex})^2 + 4\Omega^2}$
are similar in magnitude.
This large ratio validates our simplification of the 
electronic to polaritonic current conversion process into two steps occurring sequentially in space.   

\section{Condensate Momentum in the Superfluid State}
When it carries a supercurrent the condensate has a non-zero phase gradient.
As explained above, over the bulk of the coherent region we expect a pure 
LP condensate with uniform density and phase gradient,
i.e. $\Phi_\text{LP}(\mathbf{r}, t)=\phi_\text{LP} e^{i(k_\text{s}x-\omega t)}$
where $k_s$ is the condensate momentum in the $x$ direction.
The condensate momentum and current density are related by 
\begin{align}
I_\text{s} = e n_\text{s} v_\text{s} \label{super_flow}
\end{align}
where $n_\text{s}=|\phi_\text{LP}|^2$ is the density
and $v_\text{s} = \hbar k_\text{s}/m_\text{LP}$ the superfluid velocity of the LP condensate.
Owing to the dynamic nature of polariton condensates, photons are emitted continuously out of the cavity, carrying precise information about the momentum space distribution of the parent polariton particles\cite{Deng2010,Houdre1994}.
The condensate momentum $k_s$ can thus be measured using angle-resolved emission spectroscopy. \cite{Wertz2010}
At zero bias, the quasi-equilibrium condensate can either be formed spontaneously
at zero momentum using a standard non-resonant pumping scheme, 
or pumped directly with a resonant laser pump at finite momentum $k_s \neq 0$.
When an electrical bias is applied, $k_s$ is shifted with respect to the unbiased ground state (e.g. $k_s=0$) and the 
emitted photons acquire an extra transverse momentum $\Delta k=k_s$.

Quasiparticle-condensate transfer can be observed if a detectable momentum shift $\Delta k$
is achievable.  Because of their finite size, even deep in the condensate regime, polariton 
condensates have a finite width in their momentum distribution,
which places a fundamental limit on the resolution of the condensate momentum $k_s$.
It has been shown\cite{Deng0607,Mylnikov2015} that the $1/e$ width (or the FWHM) of the momentum distribution above the threshold 
pumping density is typically $0.5$ to $1 \mu m^{-1}$.
As the pumping density increases, the momentum distribution broadens slightly possibly due to 
polariton-polariton interactions.
Taking the minimum detectable $\Delta k$ interval as $\Delta k^\text{min} \sim 1 \mu m^{-1}$ and 
using Eq.(\ref{super_flow}),
we estimate that the corresponding LP supercurrent density for $k_s=\Delta k^\text{min}$ is,
\begin{align}
I_\text{s}^\text{min} \approx 1.6 \times 10^{-1} \mu A\cdot\mu m^{-1}
\end{align}
where we have taken $m_\text{LP}=10^{-4}m_0$ and $n_\text{LP}= 1\mu m^{-2}$, which is slightly larger than the threshold density.
We have marked the bias voltage at which $I_\text{s}^\text{min}$ is reached
on the $I_\text{s}-V$ curve in Fig.~\ref{Fig:IV}.
The required voltage is well in the sub-gap regime in which the transport is mainly 
due to the tail of the Fermi distribution.  

\section{Summary and Discussion}

In this article we have described a mechanism by which an electrical bias voltage applied across a unipolar 
semiconductor quantum well can drive an exciton or polariton supercurrent, and illustrated using a simplified 
model calculation that assumes ballistic transport and abrupt boundaries between a cavity region in which a 
polariton condensate is established.  Our mechanism is based on spatially inhomogeneous 
mixing between conduction and valence bands in quantum wells
or two-dimensional materials when they are dressed by interactions with the
coherent exciton and photon fields of a polariton condensate.  Because of this dressing,
the mean-field Hamiltonian of the quasiparticles, unlike the microscopic Hamiltonian,
does not conserve particle number in conduction and valence bands separately.  The conservation 
law is restored by the action of the quasiparticles on the condensate, and this action is the source which 
generates an exciton supercurrent.   A similar mechanism applies to pure excitonic condensates, and 
indeed is thought to produce an excitonic supercurrent 
when a charge current flows through the strong-field quantum-Hall excitonic condensate 
state of bilayer electron systems\cite{Eisenstein2004}.
In the polariton case, unlike the quantum Hall case, the supercurrent can in principle be 
directly measured by examining the photons that leak out of the cavity.  According to a 
qualitative estimate based on the simplified model calculation, the effect is strong enough to
produce a measurable condensate momentum shift.  

The polariton steady state in the presence of an electronic bias voltage should in principle 
be determined by solving coupled mean-field equations that differ from those described in 
Ref.~\onlinecite{Fei2016} only because the fermionic quasiparticles are in contact with 
source and drain reservoirs that have different chemical potentials.   In such a calculation,
the finite momentum of the polariton would emerge naturally from the self-consistent calculation.
A similar calculation has been performed previously\cite{Su2008} for a model excitonic condensate.
In the present case we have simplified the calculation by assuming a two step process in which the 
condensate current is generated very close to the boundary of the coherent region, and that 
it is transformed by Rabi coupling into a purely lower-polariton condensate current over a longer length scale.
This simplification allows us to ignore the finite momentum of the condensate when we 
calculate the conduction and valence band contributions to the quasiparticle current,
and is justified by the large difference between the length scales of the two processes. 

The process by which a quasiparticle current driven through a polariton condensate generates 
a polariton supercurrent is partly analogous to the process by which a quasiparticle current driven through a 
normal-superconductor-normal metallic circuit generates a supercurrent in the superconducting metal.  
Andreev scattering is a process which involves change of band character of the outgoing particle.
In the case of a metal-superconductor junction, broken particle-number conservation in the 
mean-filed Hamiltonian allows both Andreev reflection (AR), a process in which an incident
electron is reflected as a hole on the normal-metal side of the junction, and 
Andreev transmission (AT), a process in which an incident electron 
is transmitted into the superconductor with partial hole-like character.
Because the contacts are n-type, the model we have studied for polariton condensation does not 
allow the analog of AR or 
AT, but does allow a normal transmission channel involving the dressed conduction quasiparticle band.
The violation of particle-number conservation is weaker 
in the polariton case because the dressed conduction bands are still dominantly conduction band in character, 
but the effect that launches a Cooper-pair supercurrent in a superconductor is still similar to the 
effect discussed here.   

The simplified model we have used to illustrate our quasiparticle-condensate 
effect assumes ballistic scattering.  Even in samples that are completely free of 
disorder, this assumption is not realistic.  For example, the high-energy electrons emitted from the 
source can scatter by emitting phonons.  Even more important, the scenario we are imagine here 
requires some separate process, whether optical or electronic, that creates a population of 
conduction band electrons and valence band holes that supports the polariton condensate 
steady state.  Electronic quasiparticles will scatter off these electrons or holes, or 
off the population of non-resonant excitons that appears when they form bound states.
Scattering is important in limiting the amount of charge current that can flow through the 
system at a given bias voltage.  Because the supercurrent is generated immediately upon
entry of the quasiparticles into the coherent region, however, we do not expect that the amount of 
supercurrent generated for a given charge current will differ greatly from the estimate we 
have obtained.

Quasiparticle-condensate transfer can be used to deflect polariton supercurrents providing,
for example the possibility of electrical control of polariton flow at junctions.  Because Rabi coupling converts the 
condensate current to a lower polariton current, and the lower polariton state is in general spread across a 
number of quantum wells or two-dimensional material layers, the dressed quasiparticle Hamiltonian in one layer
can be influenced by bias voltages applied to another layer separated by some number of photon wavelengths.
We anticipate that the effect we have described might therefore enable interesting possibilities for 
coherent electrical coupling between two-dimensional electron systems separated by micron length scales.

This work was supported by the Army Research Office under Award No. W911NF-15-1-0466 
and by the Welch Foundation under Grant No. F1473. 

\appendix
\section{Self-consistent microscopic theory of quasiparticle-condensate transfer}
In this Appendix, we provide a microscopic explanation of supercurrent generation due to 
quasiparticle-condensate transfer based on the time-dependent Hartree-Fock (TDHF) formalism.
We focus on a purely excitonic case for simplicity.  
The polariton case can be derived similarly by including the photon field and its 
coupling to fermionic degree of freedom.

The BCS-like mean field Hamiltonian employed in our calculation breaks the global $U(1)$ symmetry
under which the fermionic field operator transforms as
\begin{align}
	\hat{\Psi} \rightarrow \hat{\Psi}' = e^{i \theta \sigma_z} \hat{\Psi}.
\end{align}
where $\hat{\Psi}=(\hat{\Psi}_c,\hat{\Psi}_v)^T$ and $\sigma$ is the Pauli matrix in the two-band basis.
Because of this spontaneous symmetry breaking, the continuity equation associated with the electron-hole density 
$\hat{\rho}_{\text{e-h}}=\hat{\Psi}^\dagger(\bm{r})\sigma_z\hat{\Psi}(\bm{r})$ and the counter flow current 
$\hat{\bm{j}}_{\text{e-h}}=\big[\hat{\Psi}^\dagger(\bm{r}) \big(\bm{p}\hat{\Psi}(\bm{r})\big)
-\big(\bm{p}\hat{\Psi}^\dagger(\bm{r})\big)\hat{\Psi}(\bm{r})\big]/(2m)$ is no longer satisfied.
Because the full Hamiltonian of the system respects this symmetry, we expect the continuity equation to hold.
In the steady state case considered in Sec.~\ref{conversion}, the conservation of the counter-flow current is restored by assuming 
that a finite condensate supercurrent is induced which cancels the non-conserving part of $\bm{j}_-$ exactly.
As pointed out in Sec.~\ref{conversion}, in the heavily studied analogous inhomogeneous 
magnetic systems the source(drain) term acts as a torque and is 
referred to as the spin-transfer torque.

In the following, we show that this conservation of counter-flow current is naturally satisfied in a self-consistent manner in the TDHF approach. 
The quasiparticle-condensate transfer effect can be illustrated by mapping to a metallic system with ferromagnetic order.

Define
\begin{align}
	\rho_{\alpha\beta}(\bm{r},\bm{r}';t) = \big\langle \hat{\Psi}^\dagger_\alpha(\bm{r}, t) \hat{\Psi}_\beta(\bm{r}', t) \big\rangle
\end{align}
where $\alpha,\beta=c,v$. The equation of motion of the density matrix writes
\begin{widetext}
\begin{align}
	i\hbar\frac{\partial}{\partial t}  \rho_{\alpha\beta}(\bm{r},\bm{r}';t) = 
	\sum_{\alpha'}\int d\bm{r}''\big[ h_{\alpha\alpha'}(\bm{r},\bm{r}'';t)\rho_{\alpha'\beta}(\bm{r}'',\bm{r}';t)
	-\rho_{\alpha\alpha'}(\bm{r},\bm{r}'';t)h_{\alpha'\beta}(\bm{r}'',\bm{r};t)\big]
\end{align}
where the full TDHF Hamiltonian is \cite{Negele1998}
\begin{align}
h_{\alpha\alpha'}(\bm{r},\bm{r}'; t) = \langle \bm{r} | \hat{T}_\alpha |\bm{r}'\rangle \delta_{\alpha\alpha'} 
+\big[
\sum_\beta\int d\bm{r}'' V(\bm{r}-\bm{r}'') \rho_{\beta\beta}(\bm{r}'',\bm{r}'';t)
\big] \delta(\bm{r}-\bm{r}') \delta_{\alpha\alpha'} -V(\bm{r}-\bm{r}') \rho_{\alpha\alpha'}(\bm{r},\bm{r}';t),
\end{align}
\end{widetext}
where  $\hat{T}_{c(v)} = \pm (\bm{p}^2/2m+E_g/2)$ is the kinetic energy operator.
We simplify our discussion by approximating the Coulomb interaction by the contact interaction $V(\bm{r})=V\delta(\bm{r})$ 
which is valid when the relevant length scale is much longer than the exciton size.
We also ignore the Hartree term of the interacting Hamiltonian since it doesn't contribute to the ``torque" effect caused by interband coherence. 
Define
\begin{align}
\rho_{\alpha\beta}(\bm{r},t) = \rho_{\alpha\beta}(\bm{r},\bm{r};t).
\end{align}
We obtain
\begin{align}
	\frac{\partial }{\partial t} \rho_{\text{e-h}}(\bm{r}, t)  =&
	- \nabla \cdot \bm{j}_{\text{e-h}}(\bm{r},t) \notag\\
    &-\frac{2}{i\hbar} 
	  \big[   \Delta(\bm{r}, t)   \rho_{cv}(\bm{r}, t) 
	          - \Delta^*(\bm{r},t) \rho_{vc}(\bm{r}, t)\big] \label{TDHF_eom}
\end{align}
where $\rho_{\text{e-h}} = \langle \hat{\rho}_{\text{e-h}} \rangle$, $\bm{j}_{\text{e-h}} = \langle \hat{\bm{j}}_{\text{e-h}} \rangle$ and we define
\begin{align}
\Delta(\bm{r}, t) = V \rho_{vc}(\bm{r}, t) . \label{sc_eq}
\end{align}
This definition can be viewed as a self-consistency equation which, when satisfied, will lead to the cancellation of the two terms in the second line of Eq.~(\ref{TDHF_eom}).
In a steady state case $\partial \rho/\partial t=0$, $\bm{j}_{\text{e-h}}$ is conserved.

The condensate-quasiparticle effect can be understood by considering the time evolution of 
a state with a bare conduction electron with finite velocity 
traveling toward an equilibrium exciton condensate at time $t=0$. 
The density matrix can be decomposed into
\begin{align}
	\rho(t) = \rho^s(t) +\rho^{qp}(t)
\end{align}
where $\rho^{s}$ and $\rho^{qp}$ corresponds to the condensate part and the quasiparticle part, respectively.
If we assume that $\rho^s$ is kept unchanged at its initial value, only $\rho^{qp}$ evolves according to the equation of motion Eq.(\ref{TDHF_eom}). This is the approximation used in Eq. (\ref{eom_c}) and (\ref{eom_v}) where a supercurrent is included {ad hoc}.
When the evolution of $\rho^s$ is included, $\rho_s$ and $\rho_{qp}$ undergo a mutual evolution in which quasiparticle-condensate transfer occurs.

To show this transfer effect more explicitly, we map the equation of motion of the exciton condensate system to that of a ferromagnetic metal by defining
\begin{align}
	\bm{s}(\bm{r},t) = \sum_{\alpha\beta} \bm{\sigma}_{\alpha\beta} \rho_{\alpha\beta}(\bm{r},t).
\end{align}

Under this mapping, Eq.~(\ref{TDHF_eom}) becomes
\begin{align}
	\frac{\partial s_z}{\partial t} = -\nabla \cdot \bm{j}_z 
	+\frac{4}{\hbar} \bm{\Delta} \times \bm{s}_\perp \label{spin_eom}
\end{align}
where $\bm{\Delta}=\text{Re} [\Delta] \bm{e}_x + \text{Im}[\Delta] \bm{e}_y$ is the effective in-plane field and $\bm{s}_\perp$ the in-plane component of $\bm{s}$. $\bm{j}_z \equiv \bm{j}_{\text{e-h}}$ is the current with spin polarized in $z$ direction.
The second term on the right-hand side of Eq.~(\ref{spin_eom}) is the well known spin-transfer torque term.
When the spin polarization of the injected current differs from the exchange field direction $\bm{\Delta}$, a mutual torque between the spin of the current and the exchange field is generated and rotates the direction of both of them around $z$ axis. 
This rotation is equivalent to a change in phase of $\Delta(\bm{r},t)$. 
The propagation of this phase perturbation corresponds to the propagation of spin-wave \cite{Rossier2004}.
In a steady state, we expect a finite phase gradient of $\Delta(\bm{r}, t)$ is generated which corresponds to a steady exciton supercurrent.


\end{document}